\documentclass[aps,twocolumn,twoside]{revtex4}

\usepackage{epsfig}
\usepackage{amsmath,amssymb,color}

\usepackage[cp1251]{inputenc}
\usepackage[english]{babel}

\parskip=\medskipamount

\newcommand{\eq}[1]{(\ref{#1})}
\newcommand{\fig}[1]{Fig.\ref{#1}}

\newcommand{\be}{\begin{equation}}
\newcommand{\ee}{\end{equation}}

\newcommand{\barr}{\begin{array}}
\newcommand{\earr}{\end{array}}

\newcommand{\beqn}{\begin{eqnarray}}
\newcommand{\eeqn}{\end{eqnarray}}

\newcommand{\bs}{\begin{subequations}}
\newcommand{\es}{\end{subequations}}

\newcommand{\bw}{\begin{widetext}}
\newcommand{\ew}{\end{widetext}}

\newcommand{\la}{\left<}
\newcommand{\ra}{\right>}

\begin{document}

\title{On the motifs distribution in random hierarchical networks}

\author{V.A. Avetisov$^1$, S.K. Nechaev$^{2,3,4}$, A.B. Shkarin$^5$}
\affiliation{$^1$N.N. Semenov Institute of Chemical Physics of the Russian Academy of Sciences,
1199911, Moscow, Russia \\ $^2$LPTMS, Universit\'e Paris Sud, 91405 Orsay Cedex, France \\
$^3$P.N. Lebedev Physical Institute of the Russian Academy of Sciences, 119991, Moscow, Russia \\
$^4$J.-V. Poncelet Labotatory, Independent University, 119002, Moscow, Russia \\ $^5$Moscow
Physical--Technical Institute, 141700, Dolgoprudnyj, Moscow district, Russia}

\date{\today}

\begin{abstract}
The distribution of motifs in random hierarchical networks defined by nonsymmetric random
block--hierarchical adjacency matrices, is constructed for the first time. According to the
classification of U. Alon {\em et al} of network superfamilies \cite{alon} by their motifs
distributions, our artificial directed random hierarchical networks falls into the superfamily of
natural networks to which the class of neuron networks belongs. This is the first example of
``handmade'' networks with the motifs distribution as in a special class of natural networks of
essential biological importance.
\end{abstract}

\maketitle

\section{Introduction and basic definitions}

Most commonly, the hierarchy of states emerges in many--particle systems of various origins with a
large number of ``frozen'' constraints with different scales, which generate multidimensional
hypersurfaces of potential energy (or free energy) with an astronomically large number of local
minima. Typical examples of such systems (often referred to as complex systems) are glasses and
globular proteins. The hierarchical concept applied to such systems presumes that local minima of
the energy landscape are clustered into hierarchically embedded basins of minima. Namely, each
large basin consists of smaller basins each of which in turn contains embedded still smaller
basins, and so on. Local minima basins are separated from one another by hierarchically ordered
barriers (the smaller the basins, the ``lower'' the barriers separating them).

Apart from the dynamic contents of the hierarchical concept, the determination of the hierarchical
organization of ``ultrametric phase spaces'' in the observed statistical regularities is of
considerable interest. A visual example of the such a structural organization is the so-called
crumpled globule discussed for the first time in \cite{gnsh}. The thermodynamically equilibrium
spatial configuration of such a globule resembles the Peano curve \cite{mand} embedded into a 3D
space. Spatial packing of a crumpled globule can be represented schematically by a single folded
motive reproduced on a growing scale. The hierarchical packing naturally leads to a
block--hierarchical network of contacts between the links of a chain described by a
block--hierarchical matrix of contacts. Naturally, the presence of inhomogeneities in the hierarchy
of crumples introduces randomness in the block--hierarchical network of contacts, which requires
the determination of statistical characteristics of an ensemble of random block--hierarchical
matrices of contacts. In recent works \cite{JStMech,JETP} we have considered statistical properties
of random hierarchical networks defined by adjacency matrices in form of block--hierarchical Parisi
matrix \cite{mezard}. Remind that a network is a set of vertices (or nodes) and connections between
them (links or edges). We suppose that in the network there are no any self--connections and
multiple edges. The network is random if any link occurs with a certain probability. The network is
directed if any link either has an orientation ($i\to j$) or is bidirectional ($i\leftrightarrow
j$). Otherwise the network is non-directed.

The investigation of statistical properties of random graphs and networks implies studying of
spectral properties of their adjacency matrices (e.g. \cite{fark,goh}), the same question can been
posed for block--hierarchical networks. It was found in \cite{JStMech,JETP} that the spectral
density of adjacency matrices has power law (``heavy'') tails, typical for scale--free networks.
This observation has been supplemented by direct investigations of such typical statistical
properties of networks as vertex degree distribution, which turned out to be abnormally wide (but
not scale--free). Hence following the conventional classification (e.g. \cite{barab}), random
hierarchical networks could be attributed to the class of scale--free (according to the spectral
density) or polyscale (according to the vertex degree distribution) networks.

Scale--free networks are associated with a variety of structures and systems, such as protein
folding and biopolymer dynamics; cell metabolism; neural, information and communication networks;
various evolutional, ecological, social and economical systems. Statistical characteristics of many
natural networks are described in the review \cite{barab}. Because of wide usage of a ``network
paradigm'' it seems quite natural that a ``handmade'' design of artificial networks with some
observed statistical characteristics close to that of natural networks is of primary importance.
Such a design might be very useful tool for searching for the correlations between the network
structural organization and the functions.

Until recently building of scale--free networks was based in almost all works on a step-by-step
growing process based on the preferential attachment method \cite{pref} and its various
modifications. In these approaches the new vertices are connected to the existing ones with
probability which depend on their current vertex degree. Most of the statistical characteristics of
artificial scale--free networks including spectral distribution of adjacency matrix were obtained
for networks built in this way. It should be noted that the preferential attachment process, which
realizes locally inhomogeneous {\em vertex} grouping, implicitly implies some mechanisms
controlling the current state of a network with long--term ``evolutionary memory''.

Unlike the preferential attachment--like methods, building of hierarchical networks is based on
constructing of hierarchically embedded clusters of {\em links}. The configuration of links is
usually described by an adjacency matrix $A$ in which for matrix elements $a_{ij}$ one has
$a_{ij}=1$ if there is a link connecting nodes $i$ and $j$ and $a_{ij}=0$ otherwise. Adjacency
matrix $A$ for non-directed graph is symmetric, i.e. $a_{ij}=a_{ji}$. To the contrary, for the
directed network $a_{ij}=1$ and $a_{ji}=0$ for the oriented link $i\to j$ and $a_{ij}=a_{ji}=1$ for
the bidirectional one $i\leftrightarrow j$. Therefore the generic adjacency matrix $A$ is not
necessarily symmetric.

It is known that the classification of many natural networks according to their vertex degree
distribution, or clustering coefficient is too rough and does not provide any relevant information
about the internal network structure. Much more detailed information about the network structure
can be provided by investigating the local topological characteristics, the so-called {\em motifs}
and their distributions \cite{alon1,alon}. For example, it is known that all networks, according to
their three--vertex oriented motifs distribution, can be divided into four {\em superfamilies}
\cite{alon}.

In this letter we announce the results of the investigation of motifs distribution in random
hierarchical networks. The key outcome consists in the fact that the motifs distribution of random
block--hierarchical networks clearly falls into one of the universal superfamilies, which includes,
in particular, networks of neurons. Besides, we claim the existence of a phase transition (in
respect to motifs' distribution) in an ensemble of block--hierarchical networks in the
thermodynamic limit.

\section{Distribution of motifs in hierarchical networks}
\label{sect:2}

\begin{figure}[ht]
\epsfig{file=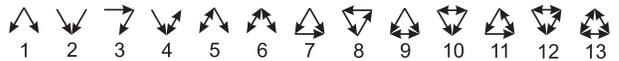, width=8cm}
\caption{Connected subgraphs--triads for directed networks.}
\label{fig:1}
\end{figure}

Remind that local topological properties of networks, both directed and non-directed, for given
number of vertices and vertex degree distribution can be characterized by the rates of connected
subgraphs. Since the number of such subgraphs grows combinatorially with their size, usually only
small subgraphs are considered. In particular, in the works \cite{alon1,alon} only subgraphs of
size 3 (triads) were analyzed for directed networks. There are 13 different configurations of such
triads. They all are enumerated in the \fig{fig:1}.

The rates of subgraphs in a given network depend on the vertex degree distribution. This
complicates the comparison of networks of different sizes and different vertex degree distributions
by the rates of their subgraphs. In order to compensate these differences, the procedure of
so-called {\em network randomization} was proposed in works \cite{alon1,alon}. In this procedure
the network experiences multiple permutations of links under the condition of conservation in each
vertex of the number of incoming, outcoming and bidirectional links. Using this method an ensemble
of randomized versions of a given network is generated, and for every subgraph the {\em statistical
significance}
\be
Z_k= {N_k-{\la N_k \ra_{\rm rand}} \over {\sigma_k}}
\label{eq:1}
\ee
is calculated, where $N_k$ is the amount of $k$-th subgraphs in the initial network and $\la N_k
\ra_{\rm rand}$ and $\sigma_k$ are correspondingly the mean and the standard deviation of $N_k$ for
the randomized networks. Subgraphs with the statistical significance essentially exceeding 1 are
called {\em motifs} \cite{alon1}. The motifs' distribution of the network under consideration is
characterized by a {\em significance profile} which is a normalized vector
\be
{\bf p}=\{p_1,..., p_m\}
\label{eq:2}
\ee
of statistical significance for all subgraphs of given size. The components of the vector ${\bf p}$
are:
\be
p_k=\frac{Z_k}{\sqrt{\sum\limits_{k=1}^m{Z_k^2}}} \qquad (k=1,...,m)
\label{eq:3}
\ee

It has been demonstrated in the papers \cite{alon1,alon} that significance profile distribution
could be used to divide networks into superfamilies. For directed networks only 4 of such
superfamilies were determined. The networks with considerably different functional properties, for
example, the neuron networks and transcriptional networks in unicellular organisms, belong to
different superfamilies. In \cite{alon1,alon} also the undirected networks were classified
according to their tetradic motifs. These networks were separated in four superfamilies as well.

\begin{figure}[ht]
\epsfig{file=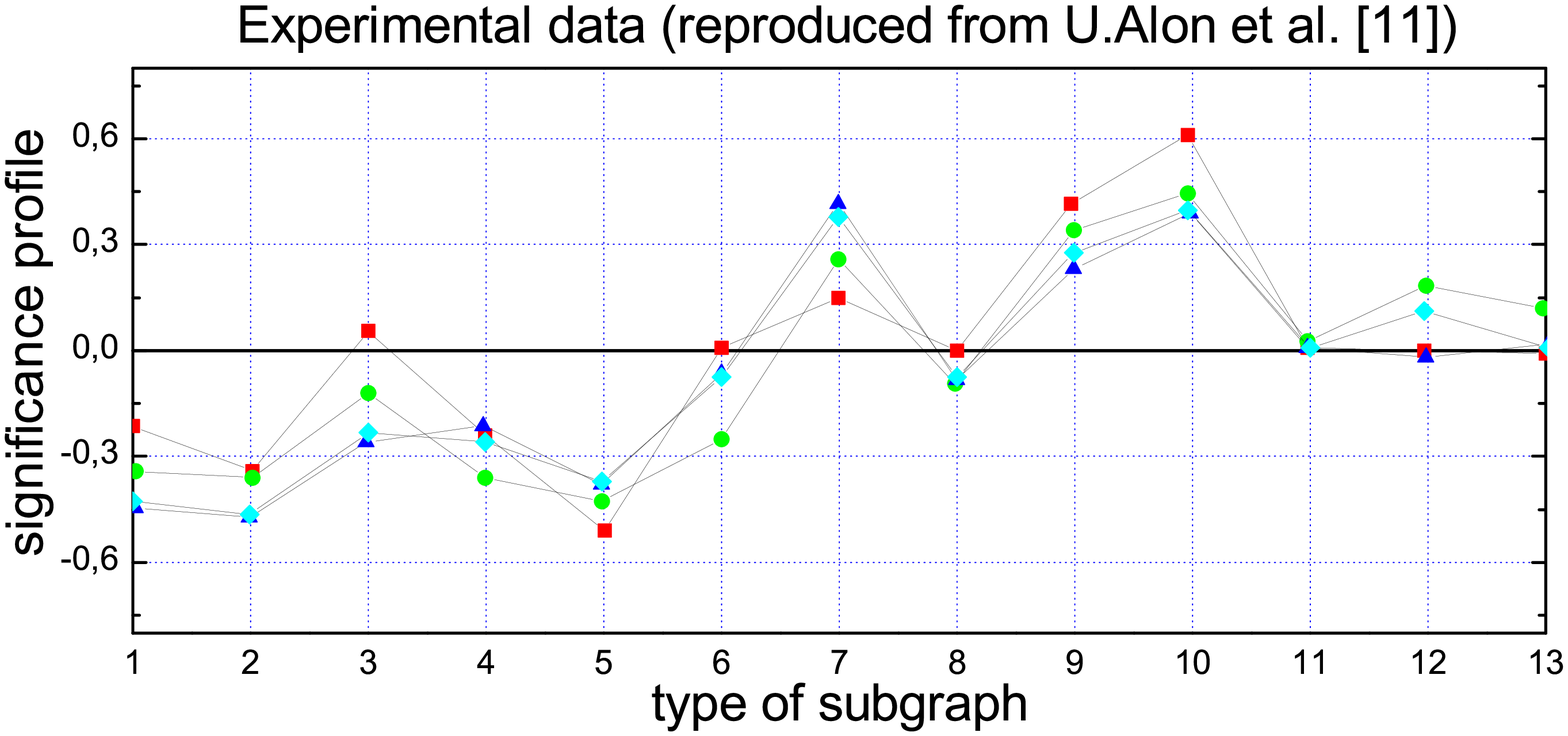, width=8.5cm} \caption{(Color online) Significance profile
distribution of motifs--triads corresponding to the superfamily of directed networks, to which the
network of synaptic contacts between the neurons in \emph{C.elegans} belong (reproduced from
\cite{alon}).}
\label{fig:2}
\end{figure}

It is interesting that artificial random scale--free networks generated by the preferential
attachment method form a separate superfamily which does not coincide with any superfamily of real
networks. In the \fig{fig:2} we have reproduced from \cite{alon} the significance profiles of one
of the superfamilies for directed networks, which later on will be compared with our results on the
distribution of motifs in the block--hierarchical random networks.

\begin{figure}[ht]
\epsfig{file=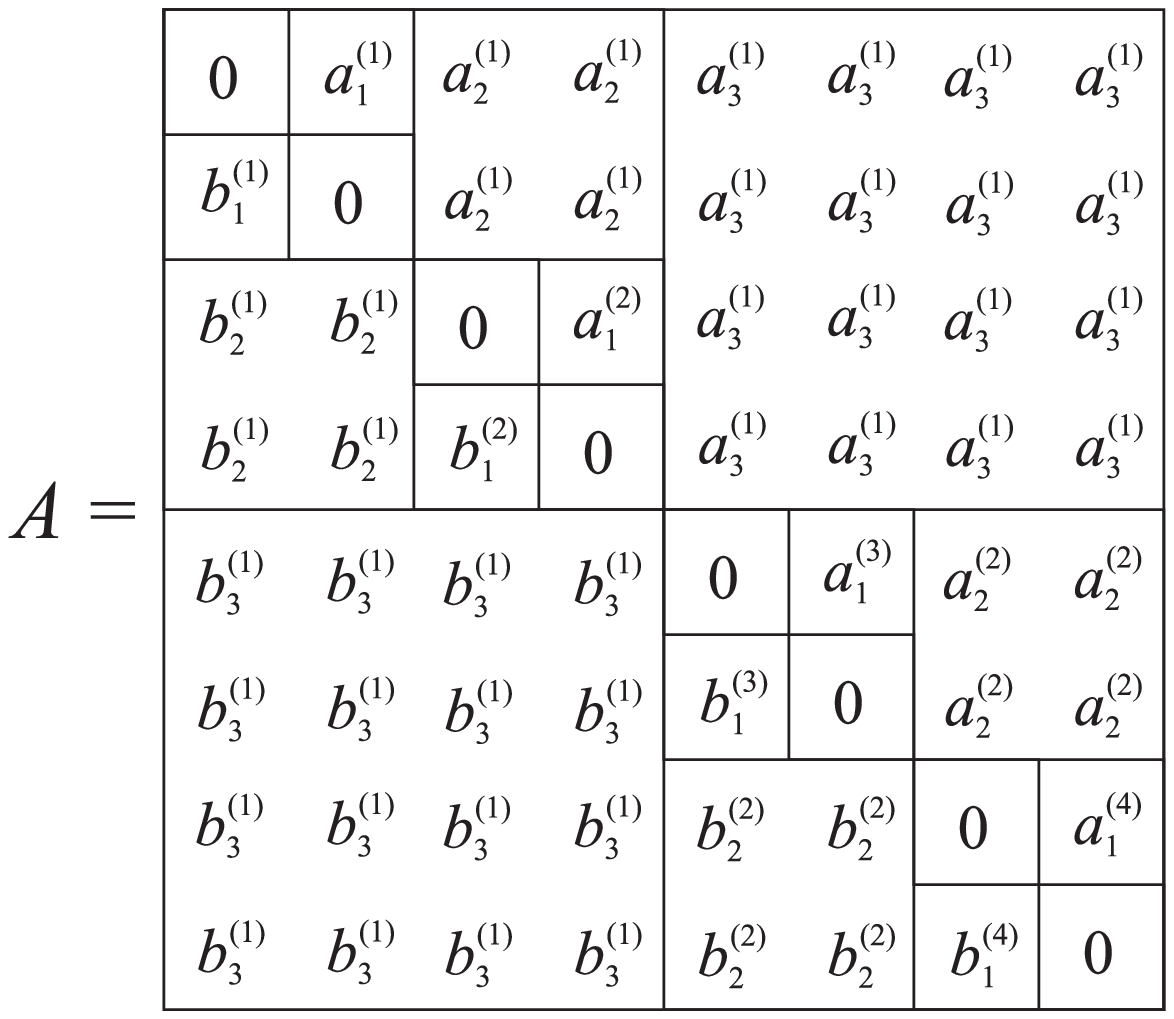, width=7cm}
\caption{Random $p$--adic ($p=2$) block--hierarchical adjacency matrix.}
\label{fig:3}
\end{figure}

The generic procedure of the random block hierarchical (RBH) network construction is as follows.
Taking $N$ points as potential vertices of our forthcoming network, we raise a hierarchical network
by connecting the vertices by edges in a specific way. We consider the adjacency matrix in form of
a $p$--adic translation--noninvariant Parisi matrix $A$. This matrix is shown in \fig{fig:3} for
$p=2$.

Since we are aimed to describe {\em directed} networks, the matrix of $A$ may not be symmetric:
$a_{ij} \neq a_{ji}$. All matrix elements, $a_{ij} \equiv a_{\gamma}^{(n)}$ and $a_{ji} \equiv
b_{\gamma}^{(n)}$, are the Bernoulli distributed random variables:
\be
\left\{a_{\gamma}^{(n)}, b_{\gamma}^{(n)}\right\} = \begin{cases} 1 & \mbox{with the probability
$q_{\gamma}$} \\ 0 & \mbox{with the probability $1-q_{\gamma}$} \end{cases}
\label{eq:4}
\ee
where $\gamma$ counts the hierarchy levels ($1\le \gamma \le \gamma_{\rm max}\equiv \Gamma$) and
$n$ enumerates different blocks corresponding to a given hierarchy level $\gamma$ (see
\fig{fig:3}). Note that the probability $q_{\gamma}$ does not depend on $n$. The full ensemble of
$N\times N$ matrices $A$, where $N=p^{\Gamma}$ ($p=2$), is completely determined by the set of
probabilities, $\{Q\}=\{q_1, q_2,..., q_{\Gamma}\}$. Thus, the elements of $A$, being the random
variables, are hierarchically organized {\em in probabilities}. In case of directed networks the
matrix elements above and below diagonal were generated independently. Below we consider the set of
probabilities, $\{Q\}$, with
\be
q_{\gamma}=p^{-\mu \gamma} \quad (\mu>0)
\label{eq:5}
\ee
In general $p\ge 2$ (we consider the case $p=2$), $\gamma=1,2,...,\gamma_{\rm max}$ is the
hierarchy level, and $\mu>0$ is a parameter. This methods allows to rise scale--free networks with
two important features. First, formation of clusters of links on each hierarchy level $\gamma$ is
uncorrelated. Second, random subgraphs associated with different hierarchy levels could be
different, so the whole network is not necessarily homogeneous.

The systematic study of statistical properties of ensembles of random graphs (networks) deals with
the investigation of the spectral properties of a graph adjacency matrix \cite{fark,goh}. Let
$\lambda_i$ ($1\le i \le N$) be the eigenvalue of the adjacency matrix. The spectral density of the
ensemble of random symmetric adjacency matrices is defined in the standard way,
\be
\rho(\lambda)=\frac{1}{N}\sum_{i=1}^N \la \delta(\lambda-\lambda_i)\ra_{\{q_1, q_2,...,
q_{\Gamma}\}}
\label{eq:6}
\ee
where $\la...\ra_{\{q_1,q_2,...q_n \}}$ denotes the averaging over the distributions of the matrix
elements of $A$. Computing numerically the spectral density, $\rho(\lambda)$, of networks with {\it
symmetric} block--hierarchical adjacency matrices, we found that the tails of the spectral density
$\rho(\lambda)$ follow a power--law (``heavy tail'') asymptotic behavior
\be
\rho(\lambda)\sim |\lambda|^{-\chi}
\label{eq:7}
\ee
In \cite{JStMech} we have found that for $\mu \in ]0,1[$ the exponents $\chi(\mu)$ takes the values
slightly below $\chi=2$. From this point of view the random hierarchical networks with symmetric
adjacency matrices are {\em scale--free}.

The hierarchical structure of clusters leads to a scale--free (polyscale) structure of network in a
broad range of the parameter $\mu$ (see \cite{JStMech}). This fact suggests that random
hierarchical networks (both symmetric and nonsymmetric) might serve as a model for certain families
of natural scale--free (polyscale) networks formed without a specific growing mechanism.

\section{Results}

\subsection{Distribution of motifs--triads in directed hierarchical networks}

\begin{figure}[ht]
\epsfig{file=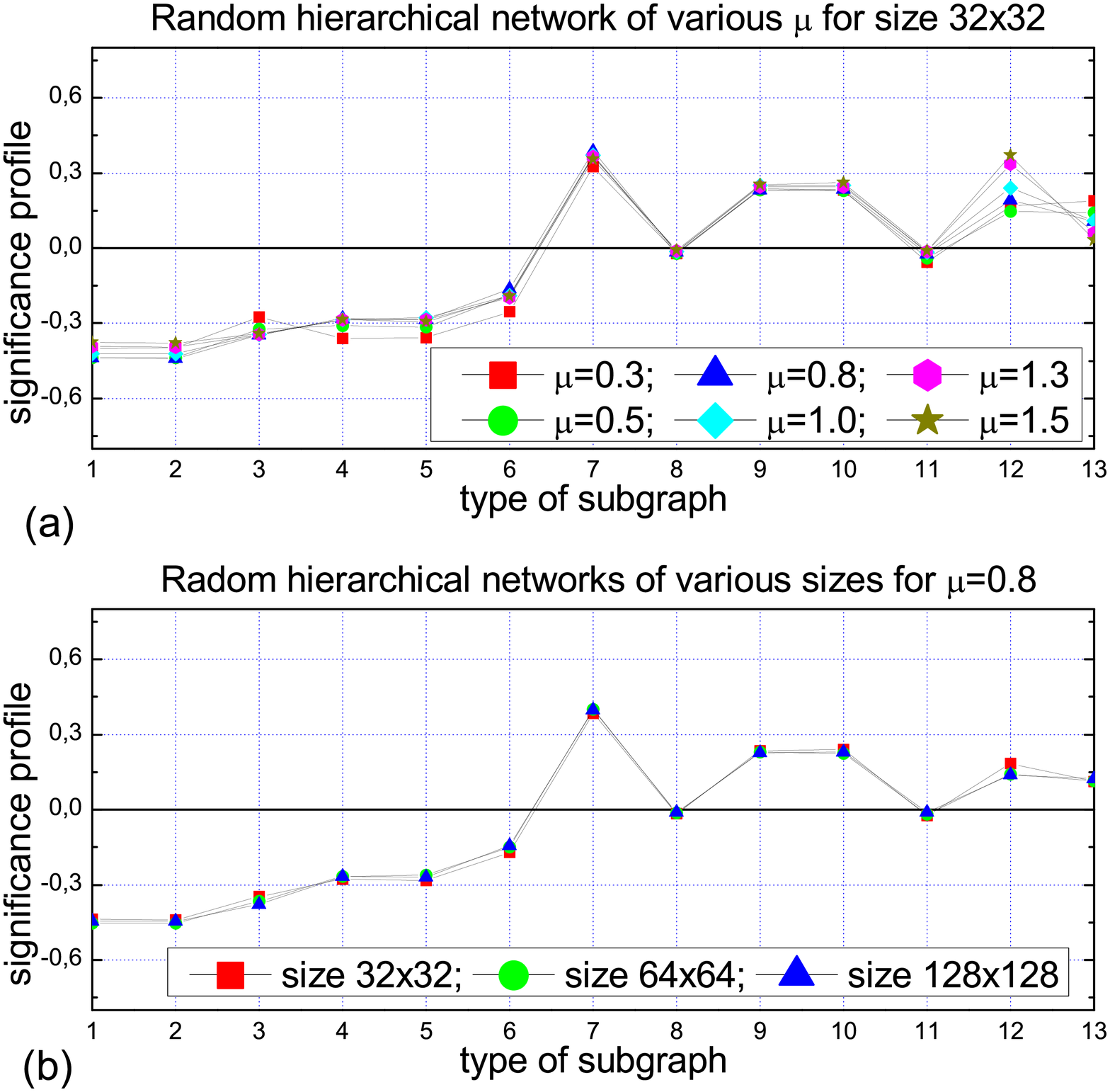, width=8.5cm}
\caption{a) (Color online) Normalized motifs distribution for hierarchical 2--adic random network of
size $32\times 32$ for different $\mu$ (see the text for details); b) Normalized motifs
distribution for hierarchical 2-adic random network of different sizes for single fixed value
$\mu=0.8$.}
\label{fig:4}
\end{figure}

The significance profile we have computed by the method described above, using triads for directed
networks. Since the networks are generated at random, for each value of $\mu$ the significance
profiles are averaged over an ensemble of corresponding random hierarchical networks. The typical
distributions of motifs for directed hierarchical networks is shown in \fig{fig:4}. As one can see,
the distributions for different values of $\mu$ look very similar and are not too sensitive
to the concrete value of $\mu$, meaning that the hierarchical networks for different $\mu$'s are
topologically similar as well.

Comparing our distribution of motifs for directed hierarchical networks depicted in \fig{fig:4}
with the ``second superfamily'' in the classification of U. Alon {\em et al} shown in \fig{fig:2}
(look \cite{alon} for more details), one sees that in a broad range of $\mu$'s our directed
hierarchical networks clearly fall into the ``second superfamily'', to which, for example, the
neuron networks belong.

We would like to emphasize that the hierarchical directed networks is, apparently, the first
example of ``hand--made'' artificial networks topologically similar to a certain superfamily of
natural networks in terms of local topological properties. We should stress that hierarchical
random networks can be built by uncorrelated generation of clusters of links, unlike the
essentially correlated preferential attachment procedure. In the light of results obtained, this
feature looks particularly interesting in the context of modelling of biological operational
systems and their evolutionary prototypes (e.g. \cite{Koonin}). From this point of view the
hierarchical networks could be of particular interest for neuron networks modelling.

\subsection{Phase transition}

We have checked our distribution of motifs (\fig{fig:4}) of directed hierarchical network on the
stability. To study this question, the following numerical experiment has been performed. First, we
have generated the block--hierarchical adjacency matrix of some directed graph, and then we
randomly spoiled this block--hierarchical structure by the following procedure.

To be precise, we scanned {\em once} all matrix elements of the adjacency matrix row--by--row from
the first element, $a_{11}$, to the last one, $a_{NN}$. Each element $a_{ij}$ we have independently
{\em switched} to the opposite value with the probability $f$, i.e. if $a_{ij}=1$, then with the
probability $f$ the element $a_{ij}$ can take the value $0$ and vis versa. Obviously, for the noise
$f=1/2$ after one run over all matrix elements, we have destroyed all hierarchical blocks and have
generated a completely random adjacency matrix corresponding to the Erd\"os--R\'enyi random graph
\cite{erdos} with an appropriate distribution of motifs. Since our motifs' distribution is measured
off the distribution of random uncorrelated graphs, the corresponding significance profile shown in
\fig{fig:4} would be definitely $0$ for $f_{\rm up}=1/2$ for all $13$ configurations of triads.

\begin{figure}[ht]
\epsfig{file=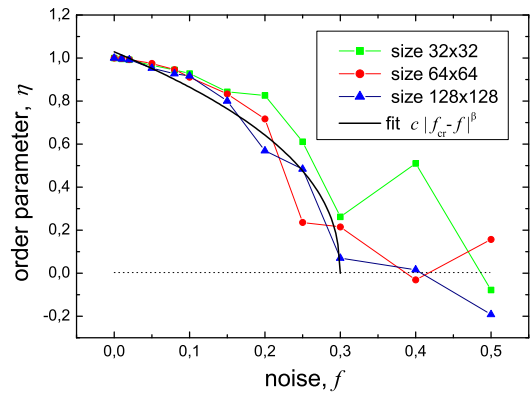,width=7cm}
\caption{(Color online) Signature of the phase transition in distribution of motifs.}
\label{fig:5}
\end{figure}

To characterize quantitatively the degree of similarity between motif's distributions for different
$f$, we define the scalar product, $\eta(f)$:
\be
\eta(f)={\bf p}(f)\, {\bf p}(0) = \sum_{k=1}^{13} p_k(f)\, p_k(0)
\label{eq:8}
\ee
where ${\bf p}(f)$ is the distribution of motifs for a given value $f$ of the noise, and ${\bf
p}(0)$ is the reference (initial) distribution of motifs for $f=0$. The value $\eta$ serves as an
``order parameter'' and its critical behavior is shown in the \fig{fig:5} for three different sizes
of networks $32\times 32$, $64\times 64$ and $128\times 128$.

Varying the ``intensity of noise'', $f$, from $f=0$ up to $f_{\rm up}=1/2$, we have noticed that
the order parameter $\eta$ becomes statistically indistinguishable from $0$ for values of $f$
significantly less than $f_{\rm up}=1/2$. Namely assuming the critical behavior
\be
\eta(f) = c\, |f_{\rm cr}-f|^{\beta}
\label{eq:9}
\ee
we have found the following numerical values $c\approx 1.72$; $f_{\rm cr}\approx 0.3$;
$\beta\approx 0.43$ for the best fit of the data corresponding the matrix of size $128\times 128$.
We can interpret this behavior as a signature of a possible phase transition characterized by the
behavior of the order parameter $\eta(f)$ in the block--hierarchical networks in the thermodynamic
limit.

In order to demonstrate that the existence of the phase transition at finite temperature is very
natural for the hierarchical system, we consider in the Appendix \label{appendix} the toy model of
the spin system on a complete graph with block--hierarchical coupling known in the literature as a
``Dyson hierarchical model'' \cite{dyson,baker}. The method used here is based on the so-called
$p$--adic Fourier transformation (see, for example \cite{VVZ} for details) and allows to re-derive the
classical results on Dyson model in few lines.

The physical meaning of the hierarchy of phase transition in the hierarchical model discussed in
the Appendix \ref{appendix} is very clear. When the temperature decreases, at first critical
temperature, $T_{\rm cr}^{(1)}$, the spins become correlated within the smallest clusters only,
where the interaction is the most strong. However, the clusters of 2nd and higher levels of
hierarchy remain still uncorrelated because the interaction of spins inside them is weaker. At the
second critical temperature, $T_{\rm cr}^{(2)}$ ($T_{\rm cr}^{(2)}<T_{\rm cr}^{(1)}$) the smallest
clusters remain correlated, together with the clusters of the 2nd hierarchical level, but the spins
of the 3rd and higher level of hierarchy remain uncorrelated, and so on. Thus, approaching zero
temperature is attended by a hierarchy of phase transition, along which larger and larger clusters
of spins become correlated. Note that the application of the spin--glass models to networks (see,
for example, \cite{Dotsenko}) suggests the opposite behavior, i.e. approaching zero temperature is
attended by extension of correlations from larger to smaller clusters of the network nodes. In this
respect, the hierarchical networks may constitute an alternative approach to modelling of natural
networks.

\section{Conclusion}

We have shown that the distribution of motifs in random hierarchical networks defined by
nonsymmetric random block--hierarchical adjacency matrices coincides with the distribution of
motifs in the second superfamily in the classification of U. Alon {\em et al} of networks
\cite{alon} to which the class of neuron networks belongs. We would like to emphasize that
apparently, this is the first example of ``handmade'' networks with the distribution of motifs as
in a special class of natural networks of essential biological importance.

Let us point out two important features of hierarchical networks constructed in our paper. First of
all, any sub-graph belonging to a particular hierarchy level is an independent graph because the
formation of clusters of links on each hierarchy level is entirely uncorrelated. Secondly, the
sub-graphs, associated with different hierarchy levels of the network, can be different, so the
network as a whole can be essentially nonuniform. In nature, the random graphs of such a
hierarchical genesis can be encountered among the networks whose origins are associated with random
events with low correlation, occurring with short evolutionary memory. The construction of such
networks in some sense is very ``simple'', ``rough'' and ``stable'' because it does not demand a
``fine tuning'' of parameters to demonstrate the desirable properties (for example, the
distribution of motifs). In particular, the networks of hierarchical genesis may by interesting as
regards prebiology or the earliest biology.

We believe that our result could shed the light on the relation between the distribution of motifs
and the structure of the adjacency matrix of a hierarchical network. However to make this relation
more profound the ``inverse'' problem should be considered as well. Namely, it would be desirable
to check if the stable distribution of motifs is uniquely related to any kind of hierarchical
organization of the network. The result of our work concerning the critical behavior of motifs'
distribution on the noise intensity may be considered as a step towards this direction. This result
demonstrates that the motifs's distribution for hierarchical network has a ``basin of stability''
below some critical value $f_{\rm cr}$ of random perturbation of the hierarchical network.

We are guided by a general conjecture that the motifs distributions corresponding to four
superfamilies of U. Alon {\em et al} could signalize the existence of islands of stability
(attractors) in a sea of possible motifs' distributions. Whether this conjecture is true or not we
hope to see in the close future.

\begin{acknowledgments}
The authors are grateful to A. Mikhailov for paying our attention to motifs distribution and to M.
Tamm and G. Kucherov for useful stimulating discussions. This work has been partially supported by
the Program No. 24 of the Presidium of the Russian Academy of Sciences and by the ERARSysBio Plus
grant $\#66$.
\end{acknowledgments}

\begin{appendix}

\section{Phase transition in hierarchical spin system in an external field}
\label{appendix}

In order to demonstrate that the existence of the phase transition is very natural in a
hierarchical system, we consider the toy model of Ising spin system with symmetric
block--hierarchical matrix of coupling constants. This model is known as ``Dyson hierarchical
model'' \cite{dyson,baker}. We outline the standard derivation of the mean--field solution of a
spin system and discuss briefly the obtained results for our specific hierarchy of coupling
constants.

Define the partition function $Z$ of one--dimensional Ising spin chain with an arbitrary matrix $U$
of coupling constants $u_{ij}$
\be
Z=\sum_{\{s_1,...,s_N\}} e^{\frac{1}{T} \sum\limits_{ij} u_{ij} s_i s_j +\sum\limits_{i=1}^N h_i
s_i}
\label{eq:10}
\ee
The spins $s_i$ ($i=1,...,N$) take the values $\pm 1$ and $h_i$ is the external field acting on the
spin $s_i$. Using the Hubbard--Stratonovich transform
\be
e^{\frac{1}{T} \sum\limits_{ij}u_{ij}s_i s_j} = \frac{\pi^{-N/2}}{\sqrt{\det U}}
\int\limits_{-\infty}^{\infty} \prod_{i=1}^N dx_i e^{-4T\sum\limits_{ij} w_{ij}x_i x_j
+\sum\limits_{i=1}^N s_i x_i}
\label{eq:11}
\ee
where $w_{ij}$ is the element $ij$ of the matrix $W=U^{-1}$ and substituting \eq{eq:11} into
\eq{eq:10}, we can integrate over all spin configurations. The partition function $Z$ reads now
\be
Z=\frac{2^N}{\sqrt{\pi^N\det U}} \int\limits_{-\infty}^{\infty} \prod_{i=1}^N dx_i e^{-T
F(x_1,...x_N)}
\label{eq:12}
\ee
where
\be
F(x_1,...x_N)=4\sum\limits_{ij} w_{ij}x_i x_j -\frac{1}{T}\sum\limits_{i=1}^N \ln \cosh(x_i+h_i)
\label{eq:13}
\ee
We evaluate \eq{eq:12}--\eq{eq:13} in the saddle--point (mean--field) approximation. The partition
function of the system is $Z=\exp\{-T F(x_1^{(0)},...,x_N^{(0)})\}$ with $F(x_1^{(0)},...,
x_N^{(0)})$ given by \eq{eq:13} with $x_i$ being the solutions of the equations
$\left.\frac{\partial F}{\partial x_i}\right|_{x_i=x_i^{(0)}} = 0$. Setting the relation between
vectors ${\bf x}^{(0)}=\{x_1^{(0)},...,x_N^{(0)}\}$ and ${\bf
y}^{(0)}=\{y_1^{(0)},...,y_N^{(0)}\}$: ${\bf y}^{(0)}=U^{-1}{\bf x}^{(0)}$, we get:
\be
4y_i^{(0)}=\frac{1}{T}\tanh\left(\sum_{j=1}^N u_{ij} y_i^{(0)}+h_i\right)
\label{eq:16}
\ee
For small arguments of $\tanh(...)$ in the r.h.s. of \eq{eq:16} one can linearize \eq{eq:16} and
rewrite it as
\be
\sum_{j=1}^N(u_{ij}-4T\delta_{ij})\,y_j^{(0)}+ h_i=0
\label{eq:17}
\ee

Suppose now that the matrix of coupling constants $U=\{u_{ij}\}$ has the block--hierarchical
structure identical to the structure of the Parisi matrix $A$ shown in \fig{fig:3}. For simplicity
we consider the matrix elements of $A$ to be nonrandom with $a_{\gamma}^{(n)}=b_{\gamma}^{(n)}=
2^{-(\alpha+1) \gamma}$, and the external field uniform $h_i=h$ for all $i\in [1,N]$.

The solution of \eq{eq:17} can be found using the methods of mathematical analysis on the field of
$p$--adic numbers $Q_p$. This technique is based on parametrization of the matrix elements
$\{u_{ij}\}$ by the pairs of rational numbers $\{z_i,z_j\}$ (see \cite{avetisov,parisi}) by such a
way that the $p$--adic norm $|z_i-z_j|_p=p^{\gamma (z_i,z_j)}$ induces block--hierarchical
structure of the matrix $U=\{u_{ij}\}$ and therefore the coupling constants $u_{ij}$ can be
represented by an appropriate (real--valued) function $u_{ij}=\rho(|z_i-z_j|_p)=a_{\gamma
(z_i,z_j)}$. In our case, $\rho(|z_i-z_j|_p)$ is chosen in the form:
\be
\rho(|z_i-z_j|_p)=\begin{cases} 0, & \mbox{$\gamma(z_i,z_j)\leq 0$} \medskip \\
2^{-(\alpha+1)\gamma(z_i,z_j)}, & \mbox{$\gamma(z_i,z_j)>0$} \end{cases}
\label{eq:21}
\ee
Now, replacing the vector ${\bf y}^{(0)}=\{y_1^{(0)},...,y_N^{(0)}\}$ by the function
$f(z_i)=y_i^{(0)}$, we can read \eq{eq:17} as follows:
\be
\sum_{j=1}^N\rho(|z_i-z_j|_p)f(z_j)-4Tf(z_i)+h=0
\label{eq:18}
\ee
Continuous analog of Eq.\eq{eq:18} in the thermodynamic limit $N\to\infty$ is
\be
\int\limits_{Q_p}\rho(|z-z'|_p)\varphi(z)\,d_pz' - 4T\varphi(z)+h=0,
\label {eq:19}
\ee
where $z\in{Q_p}$ and $d_pz$ is the Haar measure on $Q_p$. Thus Eq.\eq{eq:18} is understood as a
discrete form of the $p$--adic equation \eq{eq:19} induced by the relations $\int_{|z-z_i|_p\leq
1}\varphi(z)\,d_pz=f(z_i)$ in the coset space $Q_p/Z_p$.

The solution of Eq.\eq{eq:19} is easily found using the $p$--adic Fourier transformation (see, for
example, \cite{VVZ}). In terms of the Fourier transforms $\widetilde{\varphi}(k)$ and
$\widetilde{\rho}(|k|_p)$ of the functions $\varphi(z)$ and $\rho(|z-z'|_p)$, Eq.\eq{eq:19} looks
as
\be
\widetilde{\rho}(|k|_p)\widetilde{\varphi}(k)-4T\widetilde{\varphi}(k)+h=0
\ee
giving the solution
\be
\widetilde{\varphi}(k)=\frac{h}{4T-\widetilde{\rho}(|k|_p)}
\label{eq:20}
\ee
The function $\widetilde{\varphi}(k)$ has the poles at the set of critical temperatures $T_{\rm
cr}$, determined by the equation
$$
4T_{\rm cr}-\widetilde{\rho}(|k|_p)=0
$$
The values of $T_{\rm cr}$ can be easily determined in the closed form by knowing that
\be
\widetilde{\rho}(|k|_p)=\left\{\begin{array}{ll} 0, & |k|_p>1 \medskip \\
\Gamma_p(-\alpha)|k|_p^{\alpha} + A_{\alpha}, & |k|_p\le 1 \end{array} \right.
\label{eq:cr2}
\ee
where $\Gamma_p(-\alpha)=\frac{1-p^{-(\alpha+1)}}{1-p^{\ \alpha}}$ is the $p$--adic $\Gamma$--function
and $A_{\alpha}=(1-p^{-1})\frac{p^{-\alpha}}{1-p^{-\alpha}}$.

Since the Fourier transform \eq{eq:cr2} of the function $\rho(|z-z'|_p)$ given by
\eq{eq:21} possess discrete values $|k|_2^{\alpha}=2^{-\alpha\gamma}$, $\gamma=0,1,2,...$, the system of
hierarchically interacting Ising spins has a hierarchy of critical temperatures
\be
T_{\rm cr}^{(\gamma)}=\widetilde{\rho}(2^{-\alpha\gamma})/4
\label{eq:cr}
\ee

It should be noted that in fact the hierarchy of critical temperatures $T_{\rm cr}^{(\gamma)}$ defined by \eq{eq:cr2}-- \eq{eq:cr} exist only for $\alpha>0$. As soon as $\alpha\rightarrow 0$ the intervals between $T_{\rm cr}^{(\gamma)}$ tend to zero and the hierarchical model becomes similar to ordinary mean--field ferromagnetic system without the hierarchy of interactions.

\end{appendix}

\end{document}